\documentclass[final,12pt,times]{elsarticle}
\usepackage{amssymb}
\usepackage[]{amsmath}
\usepackage{graphics}
\usepackage{amssymb}
\usepackage{amsthm}
\usepackage{setspace}
\usepackage{epsfig}
\usepackage{subfigure}
\usepackage{booktabs}
\usepackage{epstopdf}
\biboptions{numbers,sort&compress}
\usepackage[pdftex,colorlinks,linkcolor=red,anchorcolor=blue,citecolor=green]{hyperref}
\usepackage{mathtools}
\usepackage{amsmath}
\usepackage{times}
\usepackage{anysize}
\usepackage{amsmath}
\usepackage{tikz}
\numberwithin{equation}{section}

\marginsize{2.0cm}{2.0cm}{1.0cm}{2.0cm}
\selectfont
\journal {Symmetry}
\newcommand{\sech}{{\rm \,sech}}

\begin{document}
\begin{frontmatter}

\title{Solitary waves and their interactions in the cylindrical Korteweg--de Vries equation}

\author[label2,label3]{Wencheng Hu} 
\author[label2]{Jingli Ren}%
\author[label1,label4]{Yury Stepanyants \corref{cor1}} \ead{Yury.Stepanyants@usq.edu.au}

\address[label2]{Henan Academy of Big Data/School of Mathematics and Statistics, Zhengzhou University,
Zhengzhou 450001, China;}

\address[label3]{College of Science, Zhongyuan University of Technology, Zhengzhou 450007, China;}

\cortext[cor1]{School of Mathematics, Physics and Computing, University of Southern Queensland, 487--535 West St., Toowoomba, QLD, 4350, Australia;}

\address[label1]{School of Mathematics, Physics and Computing, University of Southern Queensland, 487--535 West St., Toowoomba, QLD, 4350, Australia;}

\address[label4]{Department of Applied Mathematics, Nizhny Novgorod State Technical University, n.a. R.E. Alekseev, 24 Minin St., Nizhny Novgorod, 603950, Russia.}

\fntext[]{ORCIDs: 0000-0001-5105-8163 (W. Hu); 0000-0002-5392-291X (J. Ren); 0000-0003-4546-0310 (Y. Stepanyants)}

\begin{abstract}
We consider approximate, exact, and numerical solutions to the cylindrical Korteweg--de Vries equation. 
We show that there are different types of solitary waves and obtain the dependence of their parameters on distance. 
Then, we study the interaction of solitary waves of different types. \\
\end{abstract}


\end{frontmatter}

\newpage
\section{Introduction}
\label{Sect1}%

The study of weakly nonlinear cylindrical waves in dispersive media has a long history.
In 1959 Iordansky derived the cylindrical version of the Korteweg--de Vries (cKdV) equation \cite{Iordansky-1959} for surface waves in a fluid. 
A similar equation was later derived for water and plasma waves by various authors \cite{Maxon-1974, Ogino-1976, Miles-1978, Lipovskii-1985, Weidman-1988, Weidman-1992, Grimshaw-2019}. 
Currently, the cylindrical KdV equation is one of the basic equations of contemporary mathematical physics.
In application to the description of  outgoing waves with axisymmetric fronts, the equation in the proper physical coordinates reads:
\begin{equation}%
\label{Int001}%
\frac{\partial u}{\partial r} + \frac{1}{c}\frac{\partial u}{\partial t} - \frac{\alpha}{c} u \frac{\partial u}{\partial t} - \frac{\beta}{2c^5} \frac{\partial^3 u}{\partial t^3} + \frac{u}{2 r} = 0,
\end{equation}
where $c$ is the speed of long linear waves for which dispersion is negligible ($\beta = 0$), $\alpha$ is the nonlinear coefficient, and $\beta$ is the dispersive coefficient. 
Here $r$ stands for the radial coordinate and $t$ is time. 
The derivation of this equation is based on the assumption that the last three terms that describe the effects of weak nonlinearity, dispersion, and geometric divergence are relatively small (compared to the first two linear terms) and are of the same magnitude of smallness.
The smallness of the geometric divergence presumes that the cKdV equation is valid at big distances from the center of the polar coordinate frame where $r \gg \Lambda$, and $\Lambda$ is the characteristic width of a wave perturbation. 
A similar equation describing incoming waves can be also derived; it differs from Eq. (\ref{Int001}) only by the sign minus in front of the second term. 
In such a form the cKdV equation was used for the interpretation of physical experiments with plasma waves in laboratory chambers \cite{Maxon-1974, Hershkowitz-1974, Ogino-1976} (however, it becomes invalid when a wave approaches the origin).
The importance of the cKdV equation in water wave problems is related to circular perturbations which can appear due to ``point sources'' produced by underwater earthquakes, volcanoes, atmospheric pressure, fallen meteorites, etc. 
Besides, there are many observations when quasi-cylindrical internal waves were generated due to water intrusion in certain basins (see, for example, in the Internet numerous satellite images of internal waves generated by Atlantic water intrusions in the Mediterranian Sea).

The generalized cKdV equation was derived by McMillan and Sutherland \cite{McMillan-2010} who considered the generation and evolution of solitary waves
by intrusive gravity currents in a two-layer fluid.
Another generalised cKdV model was derived for the description of surface and internal ring waves subject to shear flows \cite{Johnson-1990, Khusnutdinova-2016a, Khusnutdinova-2016b}.
However, in this paper, we do not consider the influence of intrusions or shear flows, as well as the environment inhomogeneity on wave dynamics focusing on the structure of solitary waves and their interactions within the standard cKdV equation.

In 1976 Dryuma discovered that the cKdV equation is completely integrable \cite{Dryuma-1976} and found self-similar (but singular) solutions to this equation. 
Non-singular self-similar solutions were found later in several papers \cite{Cumberbatch-1978, Miles-1978, Nakamura-1981a, Calogero-1982}. 
There were also derived approximate solutions in the form of KdV solitons with gradually varying parameters (amplitude, width, and speed) \cite{Ko-1979, Stepanyants-1981}. 
As was shown in all these papers, amplitudes of outgoing waves decay as $A(r) \sim r^{-2/3}$, and their characteristic duration increase as $T(r) \sim r^{1/3}$.
Later exact solutions to the cKdV equation were derived by Calogero and Degasperis \cite{Calogero-1978a} (see also \cite{Calogero-1982}), as well as by Nakamura and Chen \cite{Nakamura-1981b}.
The structure of exact solutions constructed by these authors was mathematically very similar to N-soliton solutions to the KdV equation.
Despite the numerous publications on cylindrical waves described by cKdV equation, the structure of cylindrically diverging solitary waves was not been properly analysed in detail until now.
Their role in the dynamics of initial pulse-type perturbations as well as interactions with each other was not studied too. 
Therefore, the main aim of this paper is to fill in the gap in the knowledge in this field.

\section{Solitary wave solutions to the cylindrical Korteweg--de Vries equation}
\label{Sect2}%

\subsection{Dimensionless form of the cKdV equation and connection of cKdV with the plane KdV equation}
\label{Subsect2.1}%

It is convenient to study solutions of the cKdV equation in the dimensionless form.
To this end, we make the transformation:
\begin{equation}%
\label{Transform}%
r' = r, \quad \tau = -(\beta/2c^5)^{-1/3}(t - r/c), \quad v = \alpha(2c^2/\beta)^{1/3}u/6
\end{equation}
and present Eq. (\ref{Int001}) in the form (the symbol prime of $r$ can be omitted):
\begin{equation}%
\label{cKdV}%
\frac{\partial v}{\partial r} + 6 v\frac{\partial v}{\partial \tau} + \frac{\partial^3 v}{\partial \tau^3} + \frac{v}{2r} = 0.
\end{equation}

If we omit the last term in this equation, we obtain the classical KdV equation; one of its exact solutions in the form of a soliton is:
\begin{equation}%
\label{KdV-sol}%
v(r,\tau) = A\sech^2\frac{\tau - r/V}{T}.
\end{equation}
Here $A$ is the soliton amplitude, $T = \sqrt{2/A}$ is its characteristic duration, and $V = 1/(2A)$ is soliton speed. 
(Note that in this variable the speed looks a bit unusual; it is inverse proportional to the soliton amplitude $A$. However, in the original physical variables, the dimensional soliton speed is determined as $1/V_s = 1/c - (\beta/2c^5)^{1/3}(1/V) = 1/c - (\beta/2c^5)^{1/3}2A = 1/c - \alpha A_s/3c$, where $A_s$ is the dimensional soliton amplitude -- see the transformations (\ref{Transform}).
This gives $V_s = c/(1 - \alpha A_s/3) \approx c(1 + \alpha A_s/3)$, where approximation is valid for small-amplitude solitons which is in agreement with the assumption of a weak nonlinearity in the KdV equation.) Below we present an approximate and exact solutions to the cKdV equation (\ref{cKdV}).

There is a relationship between the ordinary KdV equation and cKdV equation established for the first time by A.A. and B.A. Lugovtsovs \cite{Lugovtsov-1969}, and then found also in Refs. \cite{Hirota-1979, Brugarino-1980}.
Making the transformation:
\begin{equation}
\label{LugTrans}
\tau' = -2\tau/r, \quad r' = 4/r^2, \quad v' = (v + \tau/4)/r
\end{equation}
one can reduce the classic KdV equation (Eq. (\ref{cKdV}) without the last term on the left-hand side) to the cKdV equation (\ref{cKdV}).
Formally, this allows us to get wide classes of exact solutions from the corresponding solutions of the KdV equation, including N-soliton solutions (some examples are presented in Refs. \cite{Hirota-1979, Leo-1982}).
However, all such solutions, apparently, are physically meaningless as they contain time-dependent nonuniform background.

\subsection{Asymptotic solution of the cylindrical KdV equation}
\label{Subsect2.2}%

In the cylindrical case, the soliton solution (\ref{KdV-sol}) is no longer the exact solution; however, if the last term in the cKdV equation (\ref{cKdV}) is small compared to the nonlinear and dispersive terms, then we can assume that the structure of a pulse having a shape of the KdV soliton (\ref{KdV-sol}) given at some distance $r_0 \gg \Delta \equiv VT$ remains the same in the outgoing wave, whereas its amplitude and other parameter are slowly varying function of $r$.
Therefore, the approximate solution can be presented as:
\begin{equation}%
\label{KdV-approx}%
v(r,\tau) = A(r)\sech^2\frac{\tau - \int dr/V(r)}{T(r)}.
\end{equation}

\begin{figure}[h!]
\centering
\includegraphics[width=15cm]{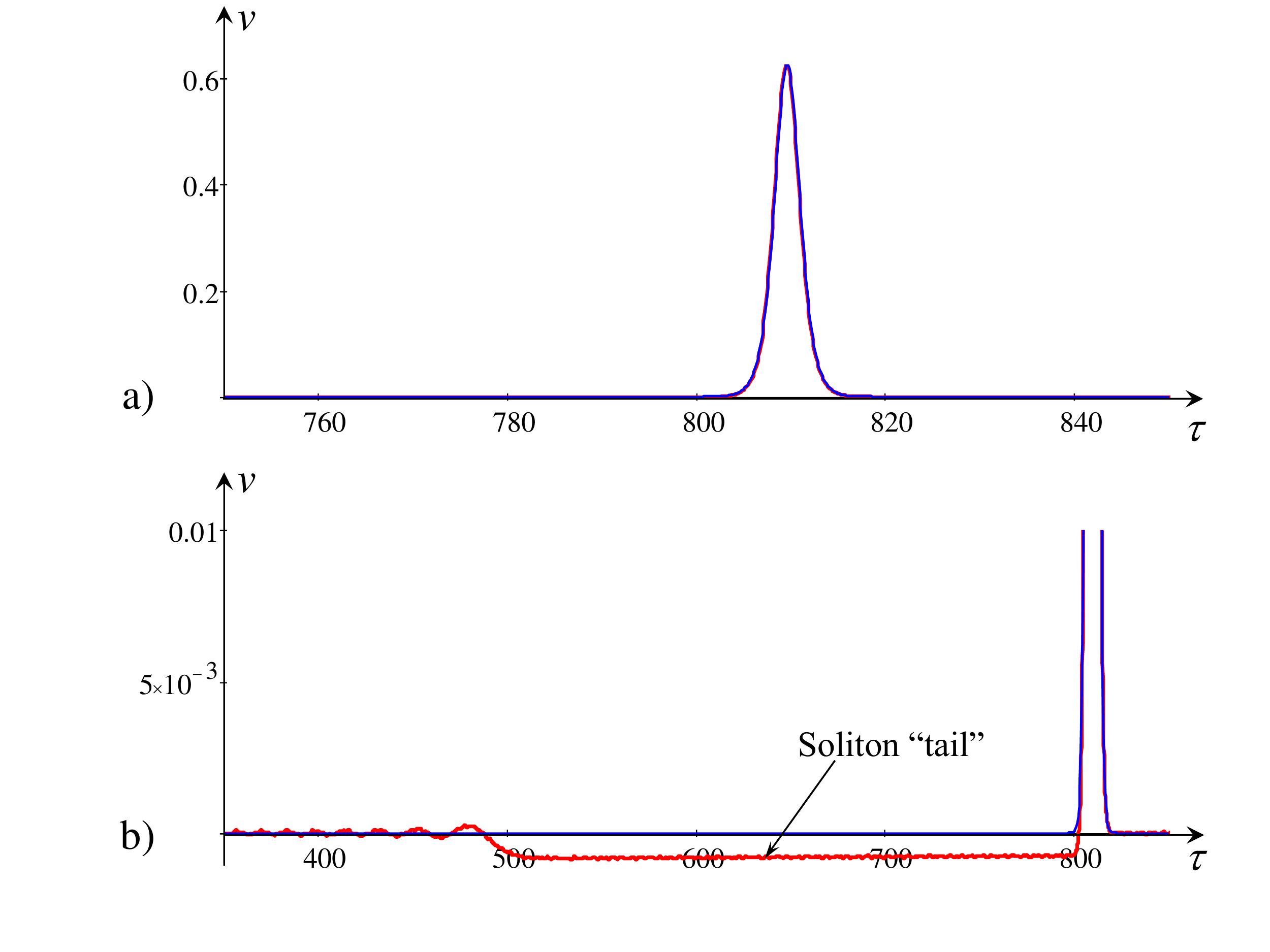}
\begin{picture}(300,6)%
\put(3,197){$0$}%
\put(3,50){$0$}%
\end{picture}
\vspace{-1.cm}
\caption{(color online).
Comparison of the approximate solution (\ref{KdV-approx}), (\ref{Adiabat}) with the numerical solution for the initial pulse in the form of a KdV soliton. Panel (a) demonstrates that the numerical solution (red line) is indistinguishable from the approximate solution (blue line). However, a small-amplitude long tail of negative polarity can be seen behind the soliton in the numerical solution when the plot is zoomed in as shown in panel (b).}%
\label{Fig01} %
\end{figure}
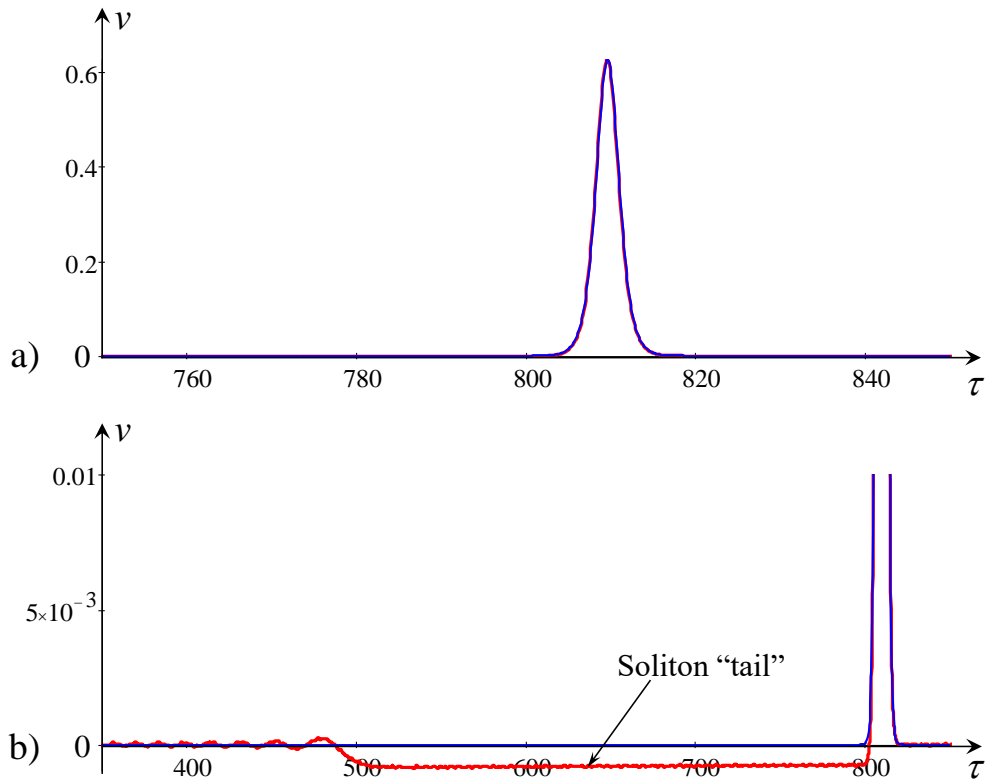

The dependence of soliton amplitude on $r$ can be found from the equation of energy flux conservation. 
Multiplying Eq. (\ref{cKdV}) by $v$ and integrating over $\tau$ from minus to plus infinity, we obtain: 
\begin{equation}%
\label{Energy-flux}%
r\int\limits_{-\infty}^{+\infty}v^2(r,\tau)\,d\tau = const.
\end{equation}
Substituting here solution (\ref{KdV-approx}) and bearing in mind the relationship between $T$ and $A$, we derive:
\begin{equation}%
\label{Adiabat}%
A(r) = A_0\left(r/r_0\right)^{-2/3}, \quad T(r) = T_0\left(r/r_0\right)^{1/3}.
\end{equation}
These are the laws of parameter variations in the nonlinear outgoing waves which were obtained in the papers cited above \cite{Ko-1979, Stepanyants-1981} and in many others (see, for example, Refs. \cite{Dorfman-1981, Fraunie-2002, Obregon-2012, Ramirez-2002}). 
Both the experimental and numerical data confirm the dependences (\ref{Adiabat}) derived in the adiabatical approximation for cylindrical solitons (see, e.g., \cite{Fraunie-2002, Ramirez-2002} and references therein). 
For the numerical study, we used the explicit finite-difference scheme described Berezin \cite{Berezin-1987} (see also \cite{Obregon-2012}).
Figure \ref{Fig01} illustrates a comparison of a typical cylindrical solitary wave as a function of $\tau$ plotted on the basis of the adiabatic formulae (\ref{KdV-approx}), (\ref{Adiabat}) and as obtained from the direct numerical solution of Eq. (\ref{cKdV}). 
The initial amplitude of the KdV soliton was chosen to be $A_0 = 1$ at $\tau = 500$ (for other amplitudes, the results were very similar).
After a while at $\tau = 809.6$, the amplitude dropped to $A(809.6) = 0.625$.
As one can see from Fig. \ref{Fig01}, the shapes of approximate and numerical solutions are not distinguishable by the naked eye.
In a more detailed comparison, one can notice that a small amplitude long tail of negative polarity forms behind the soliton in the numerical solution as shown in Fig. \ref{Fig01}b). 
The tail shape can be described in the next approximation of the asymptotic theory (see, for example, \cite{Grimshaw-1998, Ostrovsky-2015}).
The same results were obtained by Johnson \cite{Johnson-1999} who also derived the analytical expression for the tail (see also Appendix C in Ref. \cite{Grimshaw-2019} where Grimshaw estimated the decay of the tail amplitude of the negative polarity as $r^{-2/3}$).

As has been mentioned, the approximate solution is valid at a big distance from the center of a polar coordinate frame, where $r \gg \Delta$ and when the last geometric term is small compared to the nonlinear and dispersive terms. 
However, in the course of solitary wave propagation, its parameters vary and the used approximation can become invalid.
Therefore, it is of interest to estimate the validity of the approximate soliton solution (\ref{KdV-approx}), (\ref{Adiabat}) at different distances.
To this end, let us compare the last term in Eq. (\ref{cKdV}) with the nonlinear term on the soliton solution:
\begin{equation}%
\label{Compar}%
\frac{v}{2r}:6v\frac{\partial v}{\partial t} \sim \frac{T(r)}{12A(r)r} \sim \frac{T_0(r/r_0)^{1/3}}{12A_0(r/r_0)^{-2/3}r} = \frac{T_0}{12A_0} = \frac{A_0^{-3/2}}{6\sqrt{2}}.
\end{equation}
From this formula one can see that the ratio of these two terms does not depend on $r$; it remains small if it was small at the beginning when $r = r_0$. 

It is worth reminding that in this paper we study solitary waves within the framework of the cKdV equation when it is applicable to particular physical systems. 
In general, the amplitude decay of cylindrical waves can be different from the soliton amplitude dependence $A \sim r^{-2/3}$. 
As well-known, amplitudes of linear waves in cylindrical systems without dispersion vary as $A \sim r^{-1/2}$, and linear waves in cylindrical systems with dispersion vary as $A \sim r^{-1}$. 
All these amplitude dependencies for pulse-type initial perturbations were observed in experiments with electromagnetic waves in 2D lattices \cite{Stepanyants-1981, Dorfman-1981}. 
Similar results were obtained in the numerical study of radially spreading axisymmetric intrusions and solitary waves \cite{McMillan-2010}.

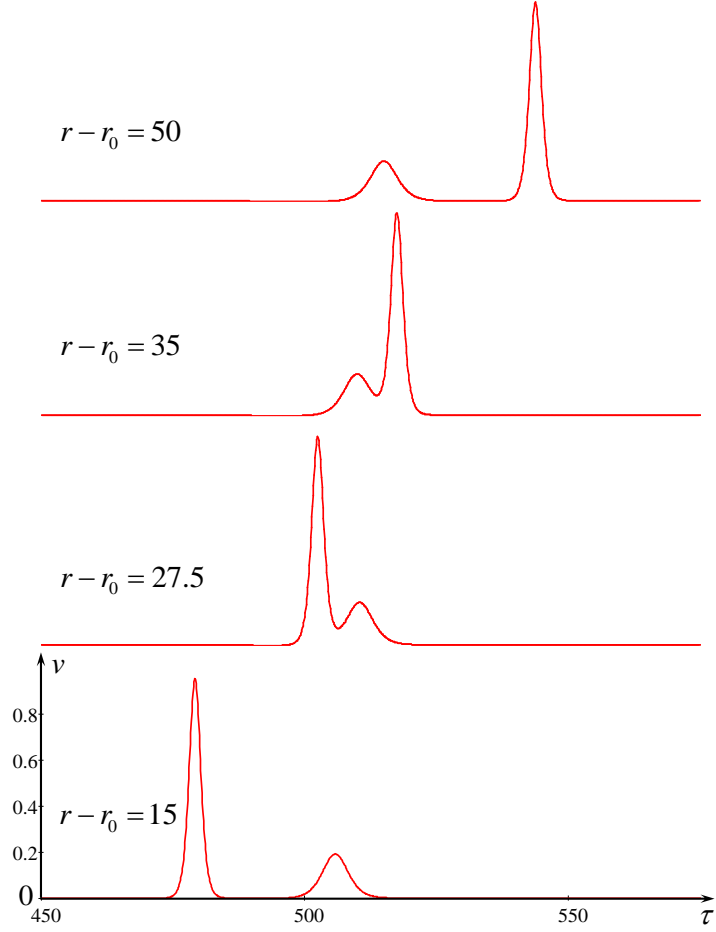
\begin{figure}[h!]
\centering
\includegraphics[width=10cm]{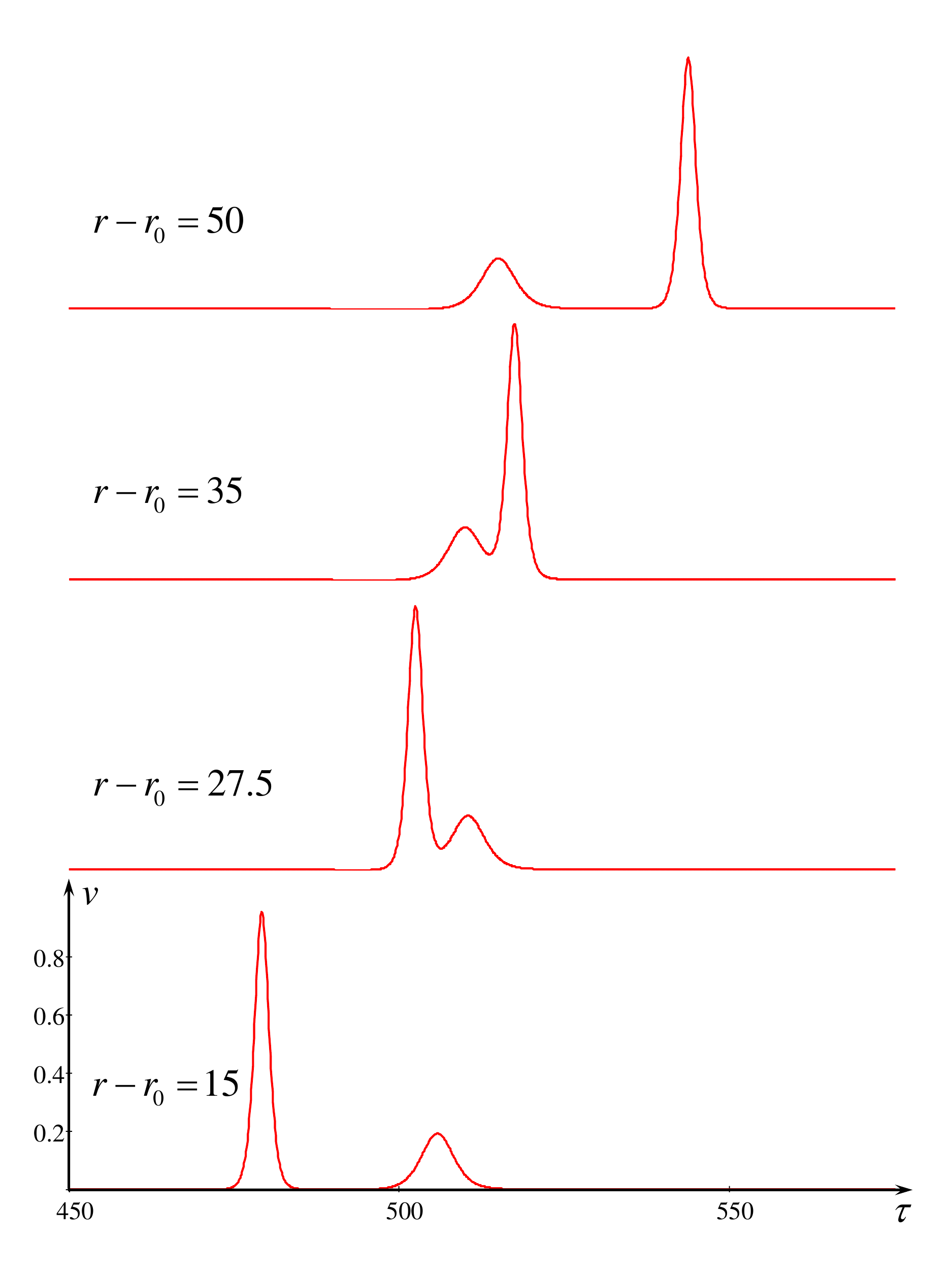}
\begin{picture}(300,6)%
\put(20,36){\small $0$}%
\end{picture}
\caption{(color online).
The typical overtaking interaction of two KdV-like solitons in outgoing cylindrical waves.}%
\label{Fig02} %
\end{figure}

Diverging KdV-like solitons interact in a similar manner as classical KdV solitons. 
Figure \ref{Fig02} illustrates the typical overtaking interaction of two KdV-like solitons within the framework of cKdV equation (\ref{cKdV}) obtained by direct numerical modeling of this equation with the initial condition in the form of two KdV solitons of different amplitudes ($A_1 = 0.2; A_2 = 1$).

\subsection{Exact solutions of the cKdV equation}
\label{Subsect2.3}%

\begin{figure}[h!]
\centering
\includegraphics[width=8.4cm]{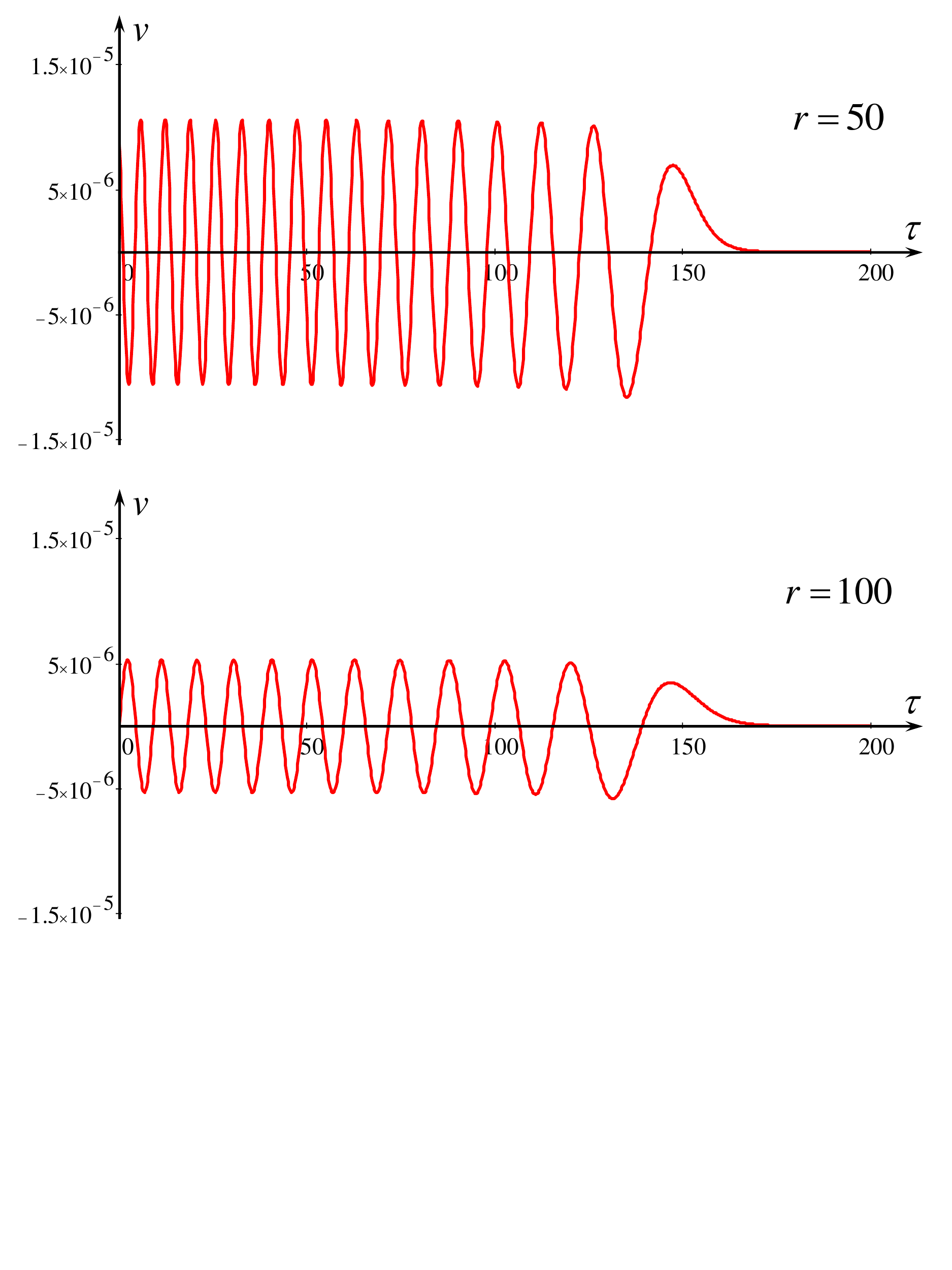} \includegraphics[width=8.4cm]{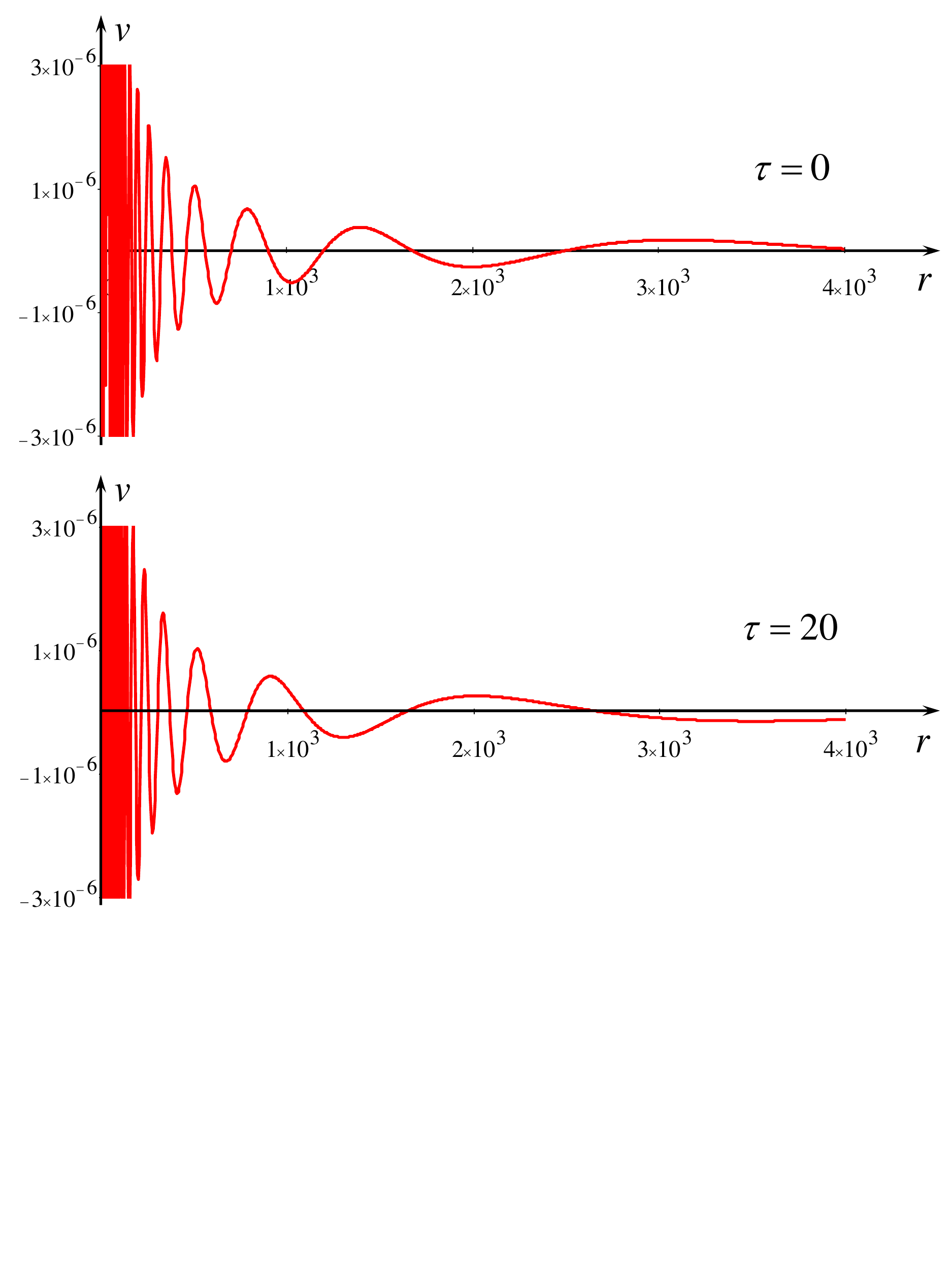}
\vspace{-2.7cm}
\caption{(color online). The typical exact solution of the cKdV equation in terms of the Airy function $\mbox{Ai}(z)$ (\ref{AAiry}) with the following parameters: $\varepsilon = -0.01$, $\rho = 1$, and $\tau_1 = 150$. In the left column, one can see the dependence of $v(\tau)$ for two distances, $r = 50$ and $r = 100$; in the right column, the solution is presented as a function of $r$ for two different times, $\tau = 0$ and $\tau = 20$. (Note that In the vicinity of the origin, the plot is simply cut; therefore, it looks that the solution is constant.)} %
\label{Fig03} %
\end{figure}

The first nontrivial exact solutions to the cKdV equation were obtained by Calogero and Degasperis \cite{Calogero-1978a}. 
Solutions were presented in terms of the Airy function $\mbox{Ai}(z)$.
As was shown later by Nakamura and Chen \cite{Nakamura-1981b}, exact solutions can be presented through the Hirota transform:
$v(r,\tau) = 2\partial^2f(r,\tau)/\partial \tau^2$.
Then, the simplest solution is:
\begin{equation}%
\label{AAiry}%
f(r,\tau) = 1 + \frac{\varepsilon \rho^2}{\left(12r\right)^{1/3}} \left\{\left[z(r,\tau) - z_1(r,\tau_1)\right]\mbox{Ai}^2(z - z_1) - \left[\mbox{Ai}'(z - z_1)\right]^2\right\},
\end{equation}
where $\varepsilon$, $\rho$, and $\tau_1$ are some arbitrary constants, and
\begin{equation}%
\label{ZandZ1}%
z(r,\tau) = \frac{\tau}{ \left(12r\right)^{1/3}}, \quad z_1(r,\tau_1) = \frac{\tau_1}{\left(12r\right)^{1/3}}.
\end{equation}
The symbol prime in Eq. (\ref{AAiry}) stands for differentiation with respect to the function argument. 
Note that in terms of the function $f(r, \tau)$, solution (\ref{AAiry}) is the typical self-similar solution on the constant pedestal. 
However, in the original variable $v(r, \tau)$, the corresponding solution is more complicated, it is neither self-similar nor a traveling-wave solution.
One of the typical exact solutions is plotted in Fig. \ref{Fig03} for the particular parameters $\varepsilon = -0.01$, $\rho = 1$, and $\tau_1 = 150$. 
This solution represents a wave that pulls into the origin as one can see from the right columns of Fig. \ref{Fig03}. 
Approaching the origin, the wavelength drastically decreases and goes to zero. 
However, in the vicinity of the origin solution becomes invalid anyway because, as mentioned above, the cKdV equation is applicable only at relatively big distances from the origin.
Apparently, such solutions are out of physical interest.

The genuine self-similar solution in terms of function $v(r, \tau)$ can be obtained if we set $\varepsilon \rho^2 \to \infty$ \cite{Johnson-1979}.
Then, we obtain:
\begin{equation}%
\label{ssAiry}%
v_{ss}(r,\tau) = \frac{2}{\left(12r\right)^{2/3}}\frac{d^2}{dz^2} \ln{\left\{\left[z(r,\tau) - z_1(r,\tau_1)\right]\mbox{Ai}^2(z - z_1) - \left[\mbox{Ai}'(z - z_1)\right]^2\right\}}.
\end{equation}
Such a solution was considered in \cite{Johnson-1980} in application to the water-wave problem.

The self-similar solution to the cKdV equation can be obtained if we seek a solution in the form $v(r, \tau) = r^\alpha F(\xi)$, where $\xi = r^\beta \tau^\gamma$ (the similar approach was used in \cite{Karpman-1975} for the KdV equation). 
Substituting this form of the solution in Eq. (\ref{cKdV}), we obtain after simple manipulation that function $F(\xi)$ must satisfy the ODE:
\begin{equation}%
\label{ssEq}%
F''' + 6FF' - \frac{1}{3}zF' + 3F = 0
\end{equation}
provided that $\alpha = -2/3$, $\beta = -1/3$, $\gamma = 1$.
This agrees with the solution (\ref{ssAiry}) if we set $F = v_{ss}\left(12r\right)^{2/3}/2$.

Calogero and Degasperis wrote that solutions that they constructed ``are in some sense the analogous of the single-soliton solutions (although they are not quite localised, having a slowly vanishing wiggling tail)''. 
The analysis of solution (\ref{AAiry}) shows that it describes a wave perturbation that decays in space as $r^{-2/3}$ whereas its duration increases with the distance as $r^{1/3}$, i.e. these quantities vary in space in the same manner as the parameters (amplitude and duration) of a solitary wave in the approximate solution (\ref{KdV-approx}), (\ref{Adiabat}).
Even more complicated solutions mathematically similar to N-soliton solutions can be constructed but all of them are far from real solitary waves.

Nakamura and Chen \cite{Nakamura-1981b} found that compact pulse-type solutions can be obtained if one replaces the first-kind Airy function $\mbox{Ai}(z)$ in the solution (\ref{AAiry}) with the second-kind Airy function $\mbox{Bi}(z)$.
Then, the simplest solution looks pretty much the same as the KdV soliton, at least in its leading part.
\begin{figure}[b!]
\centering
\includegraphics[width=15cm]{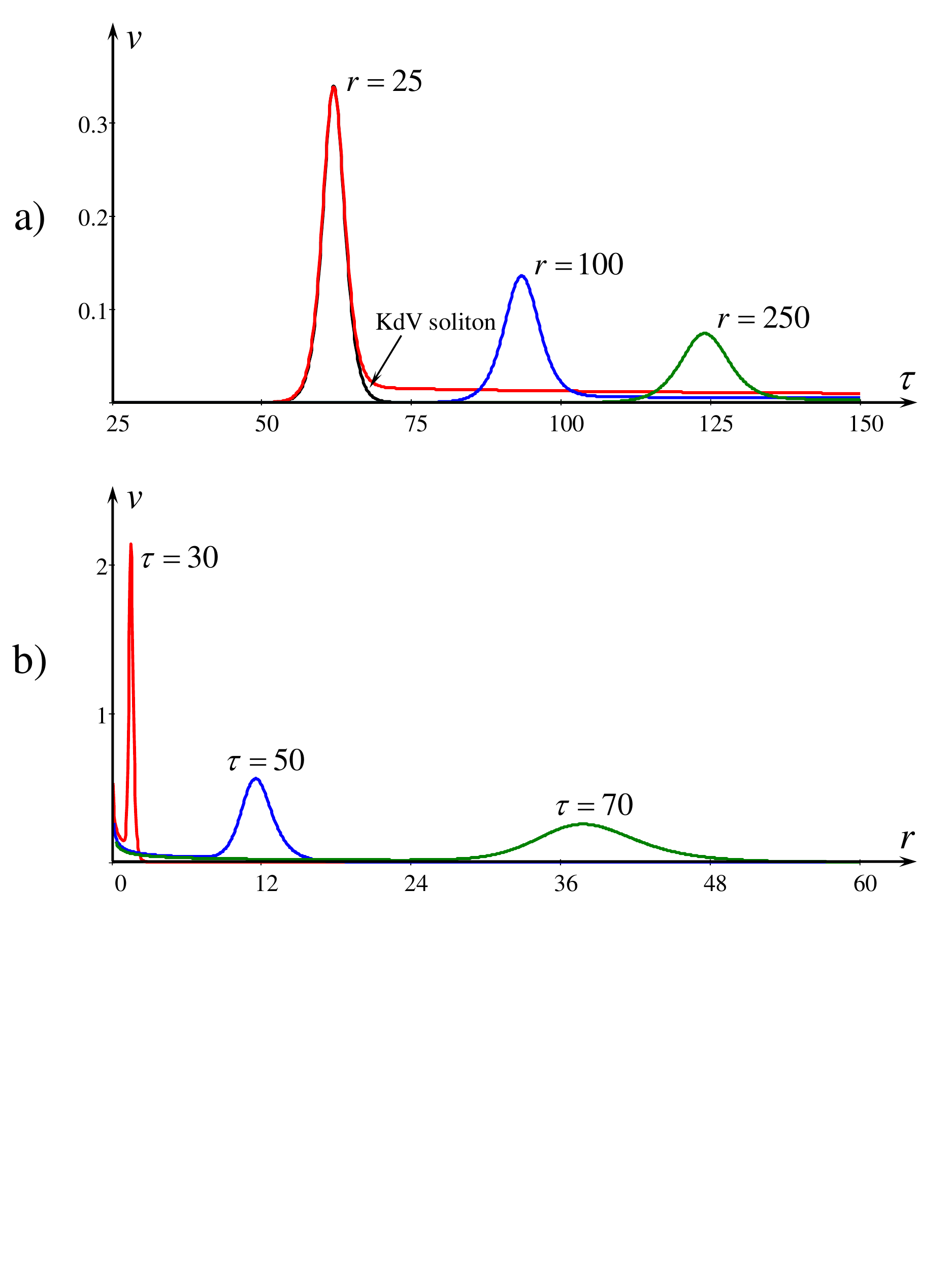}
\begin{picture}(300,6)%
\put(-22,192){$0$}%
\put(-22,400){$0$}%
\end{picture}
\vspace{-5.5cm}
\caption{(color online). Exact solution of the cKdV equation in terms of the second-kind Airy function $\mbox{Bi}(z)$ (\ref{AAiry}) with the following parameters: $\varepsilon = 10^{-4}$, $\rho = 10^{-3}$, and $\tau_1 = 10$. Panel (a) shows the dependence of the solution on time $\tau$ for the fixed distances, and panel (b) shows the dependence of the solution on distance $r$ for the fixed times.} %
\label{Fig04} %
\end{figure}
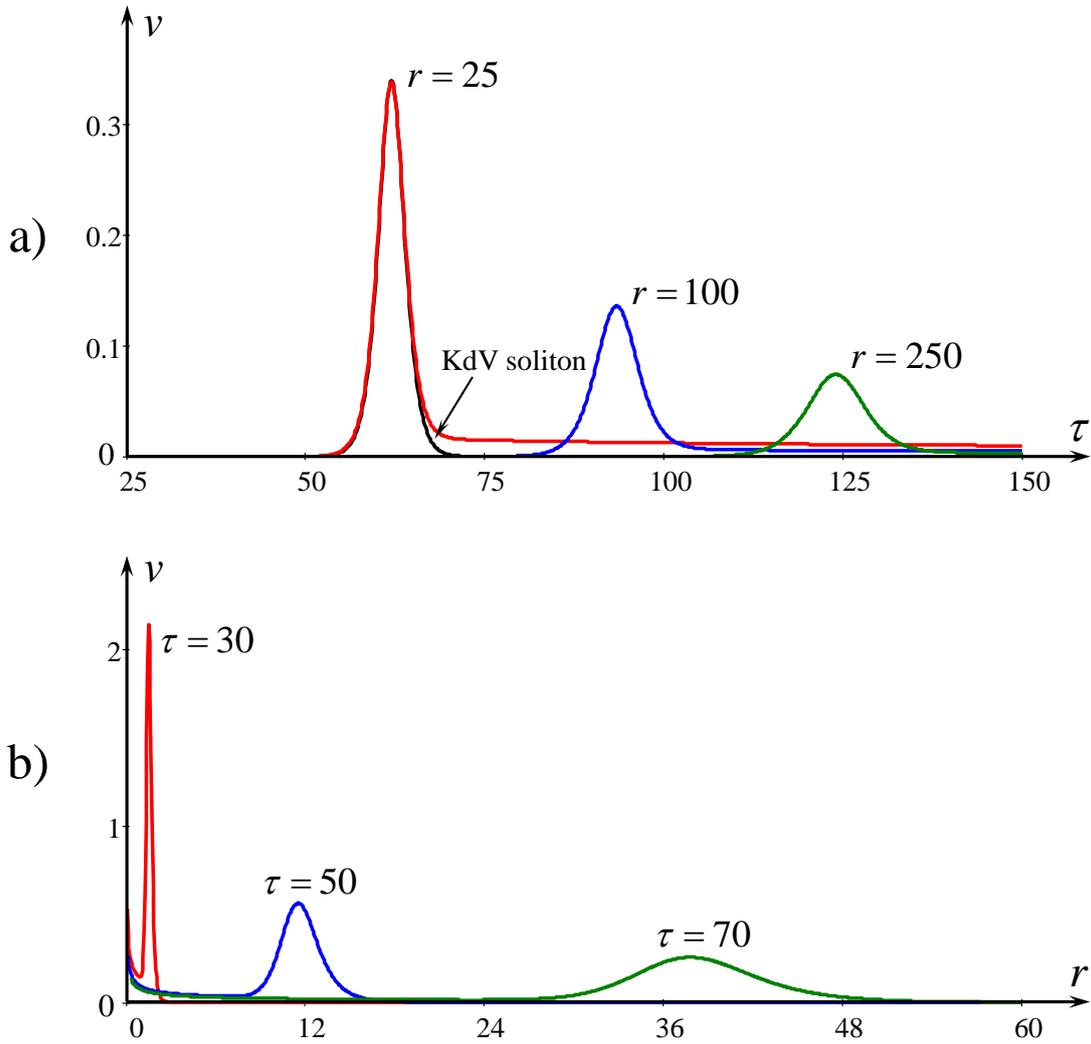
As an example, we show in Fig. \ref{Fig04}a) the comparison of solution (\ref{AAiry}) with the function $\mbox{Bi}(z)$ with the KdV soliton of the same amplitude at $r = 25$. 
As one can see, the leading parts of these solutions are practically the same; the only difference is in the rear parts of the solutions.
The same good agreements were confirmed for the solutions of equal amplitudes at other distances.
However, in contrast to KdV-like solitons, solitary waves in the solution of Nakamura and Chen \cite{Nakamura-1981b} are accompanied by well-visible positive polarity tails (cf. Fig. \ref{Fig01}b).
Solutions with Airy functions of the second-kind $Bi(z)$ are also singular at $r = 0$ like solutions with Airy functions of the first-kind $Ai(z)$ (see, for example, Fig. \ref{Fig04}b).
However, in this kind of solutions, the leading part being far from the origin, make sense and their shapes are well-approximated by KdV solitons as shown in Fig. \ref{Fig04}a).

\begin{figure}[h!]
\centering
\includegraphics[width=15cm]{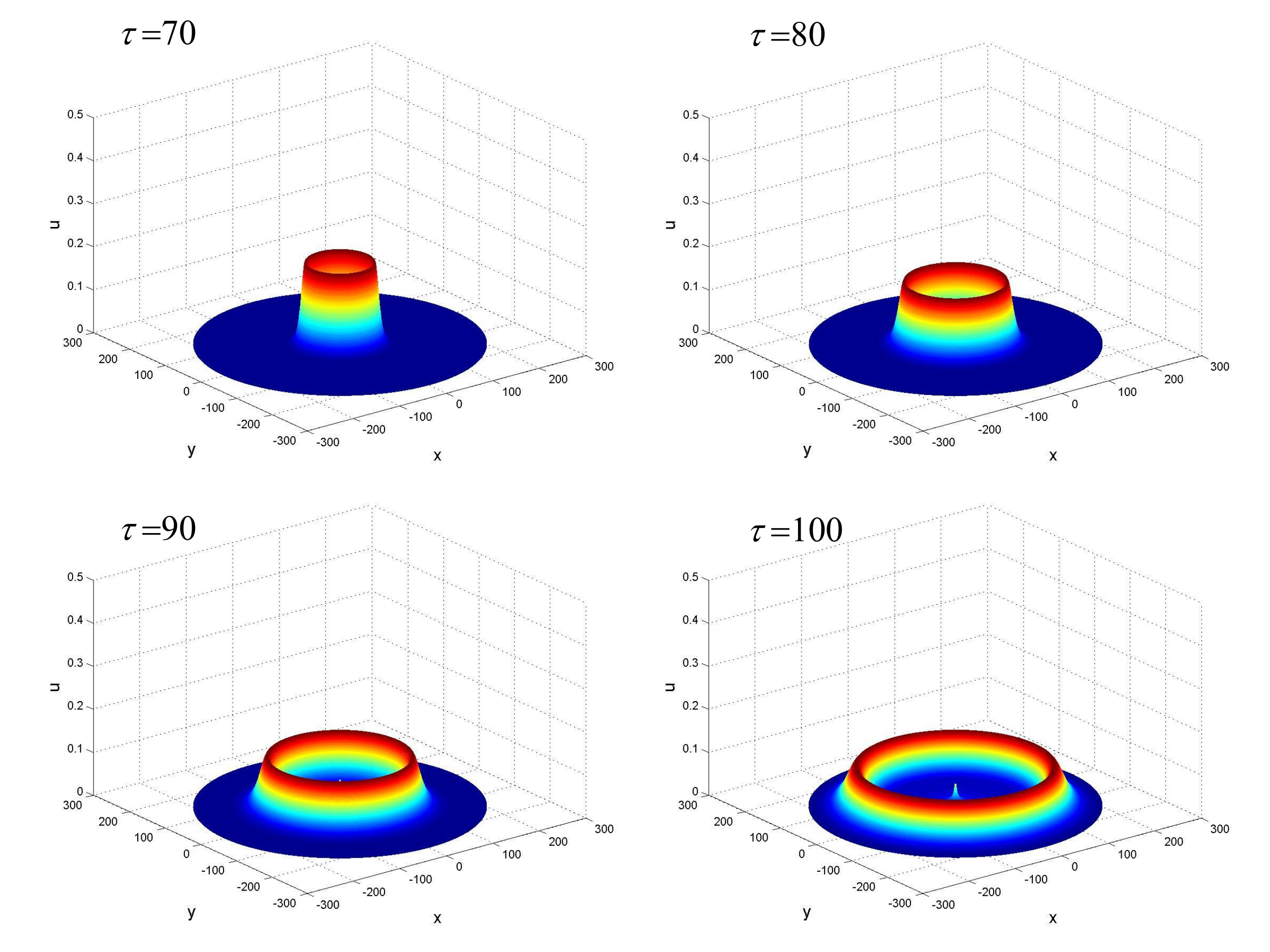}
\caption{(color online). The typical cylindrically diverging self-similar soliton is described by function (\ref{AAiry}) with the Airy function of the second kind $\mbox{Bi}(z)$ (\ref{AAiry}). The plot was generated for the same parameters as in Fig. \ref{Fig04}. Here $x$ and $y$ are the Cartesian coordinates such that $r^2 = x^2 + y^2$.} %
\label{Fig05} %
\end{figure}

Despite solutions (\ref{AAiry}) with either first-kind or second-kind Airy functions are not exactly self-similar or traveling-wave solutions, we will call, conditionally solution (\ref{AAiry}) with the second-kind Airy function $\mbox{Bi}(z)$ the {\it self-similar soliton} (ss-soliton). 
Figure \ref{Fig05} shows the diverging ss-soliton at different time moments. 
In the last frame at $\tau = 100$, one can see a singularity at the center $r = 0$.

The ``two-soliton solution'' in terms of function $f(r,\tau)$ can be presented in the form \cite{Nakamura-1981b}:
\begin{equation}
\label{eq:005}
f(r,\tau) = 1 + \varepsilon( a_{11}+a_{22})+\varepsilon^{2} \left|\begin{array}{cc}
a_{11} & a_{12}\\
a_{21} & a_{22}
\end{array}\right|=
\left|\begin{array}{cc}
1+\varepsilon a_{11} & \varepsilon a_{12}\\
\varepsilon a_{21} & 1+\varepsilon a_{22}
\end{array}\right|,
\end{equation}
where the quantities $a_{ij}$ are defined by the following expressions: 
\begin{eqnarray}
a_{ij} &=& \frac{\rho_{i}\,\rho_j}{\left(12r\right)^{1/3}} \frac{w_{i}(z - z_{i})\, w'_j(z - z_j) - w'_{i}(z - z_{i})\, w_j(z-z_j)}{z_{i}-z_j}, \quad i \ne j; \label{eq:006a} \\
a_{ii} &=& \frac{\rho_{i}^2}{\left(12r\right)^{1/3}}
\left\{\left(z - z_i\right)w_i^2(z - z_i) - \left[w_i'(z - z_i)\right]^2 \right\}, \quad i = j. \label{eq:006b}
\end{eqnarray}
where $i,j = 1, 2$, and $w_{i}(z)$ are either Airy function of the first kind $\mbox{Ai}(z)$ or Airy function of the second kind $\mbox{Bi}(z)$. 
However, as has been aforementioned, solutions with the function $\mbox{Ai}(z)$ do not represent pulse-type waves; therefore, we consider further only solutions with the second-type Airy function $\mbox{Bi}(z)$.

\begin{figure}[b!]
\centering
\includegraphics[width=16cm]
{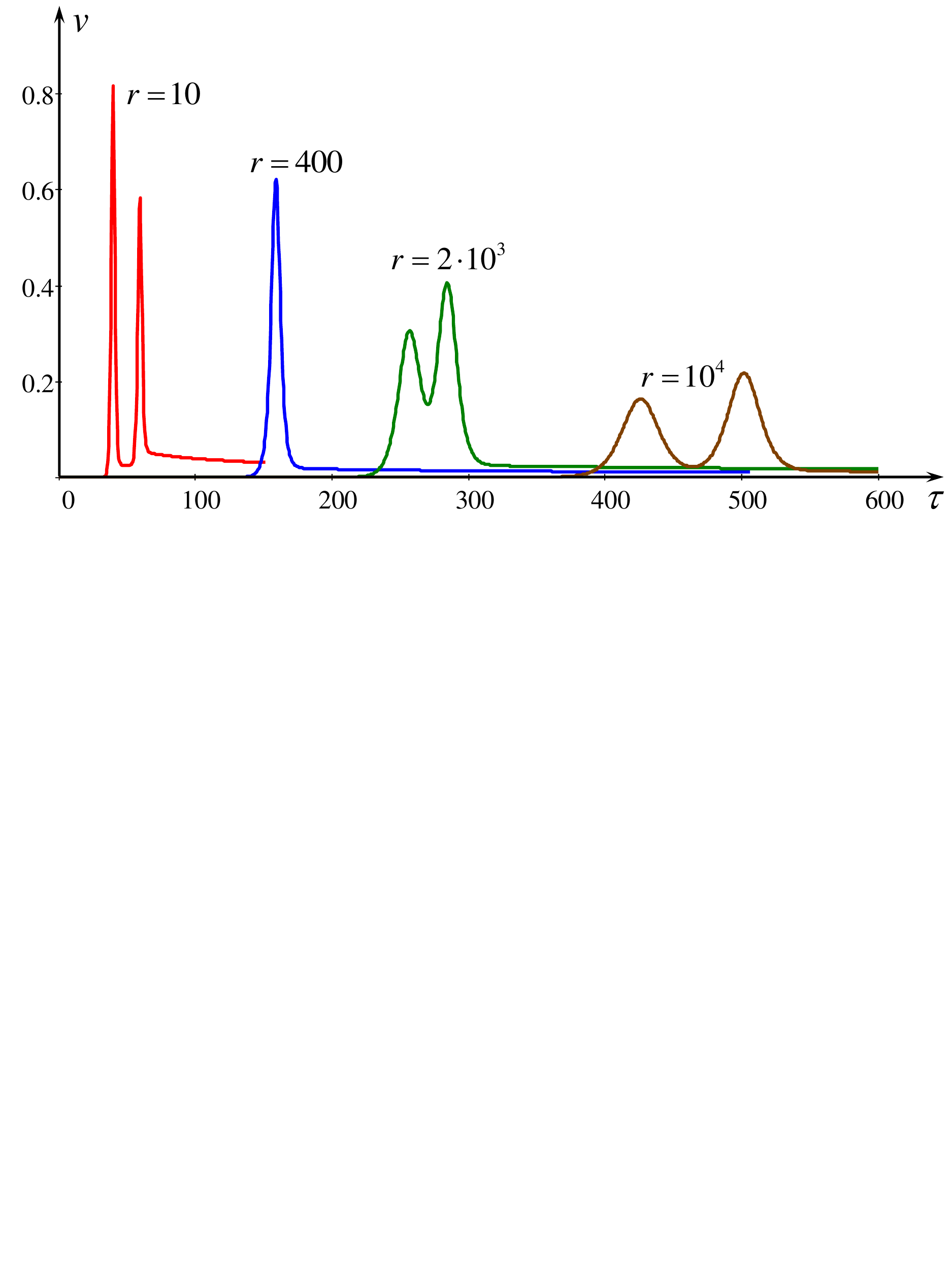}
\begin{picture}(300,6)%
\put(-58,390){$0$}%
\end{picture}
\vspace{-12.5cm}
\caption{(color online). Exact two-soliton solution of the cKdV equation in terms of the second-kind Airy function $\mbox{Bi}(z)$ as per Eqs. (\ref{eq:005})--(\ref{eq:006b}) with the following parameters: $\varepsilon = 10^{-4}$, $\rho_1 = 10^{-3}$, $\rho_2 = 10^{-6}$,  $\tau_1 = 25$, and $\tau_2 = -10$. 
To make graphics clearly visible, we multiplied function $v(\tau)$ by 4 at $r = 400$, by 16 at $r = 2\cdot 10^3$, and by 25 at $r = 10^4$.} %
\label{Fig06} %
\end{figure}
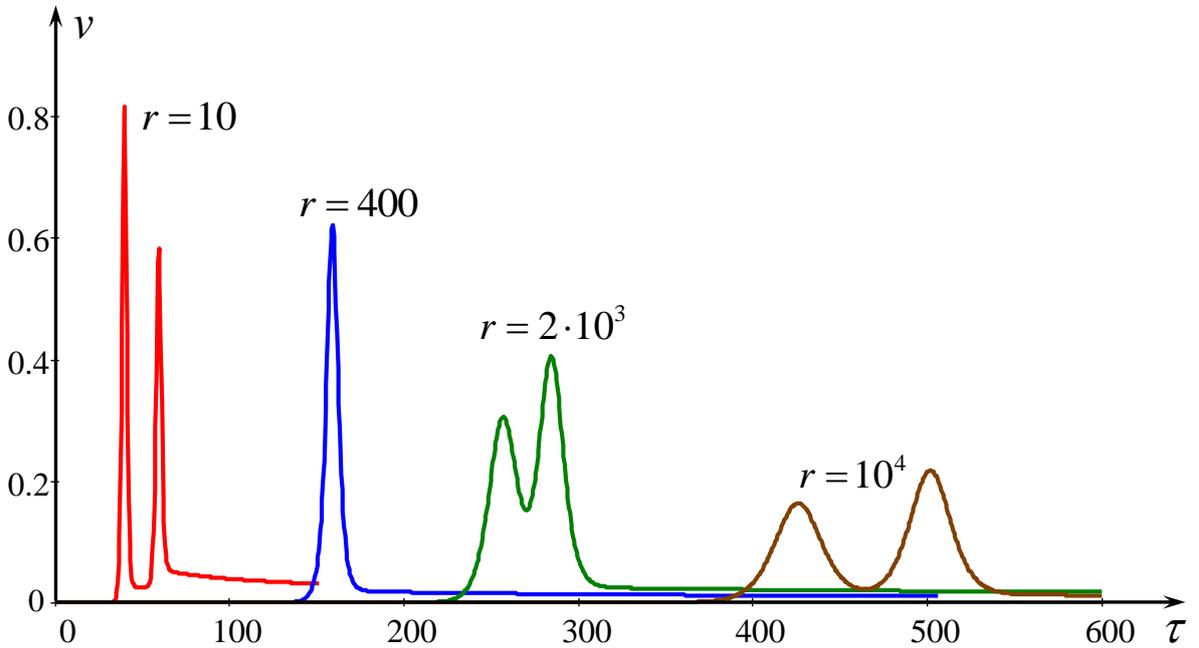

A typical two-soliton solution described by Eqs. (\ref{eq:005})--(\ref{eq:006b}) with $w(z) \equiv \mbox{Bi}(z)$ is illustrated by Fig. \ref{Fig06}. 
In this figure, one can see the time dependence of function $v(\tau)$ at three distances from the center.
The interaction of two ss-solitons resembles the overtaking type interaction of KdV solitons \cite{Gorshkov-2010} when two peaks merge at some distance (at $r = 400$ in our figure) and then, they slowly separate.
However, the separation lasts a very long time and even at big distances the pulses remain coupled as illustrated by Fig. \ref{Fig06}.

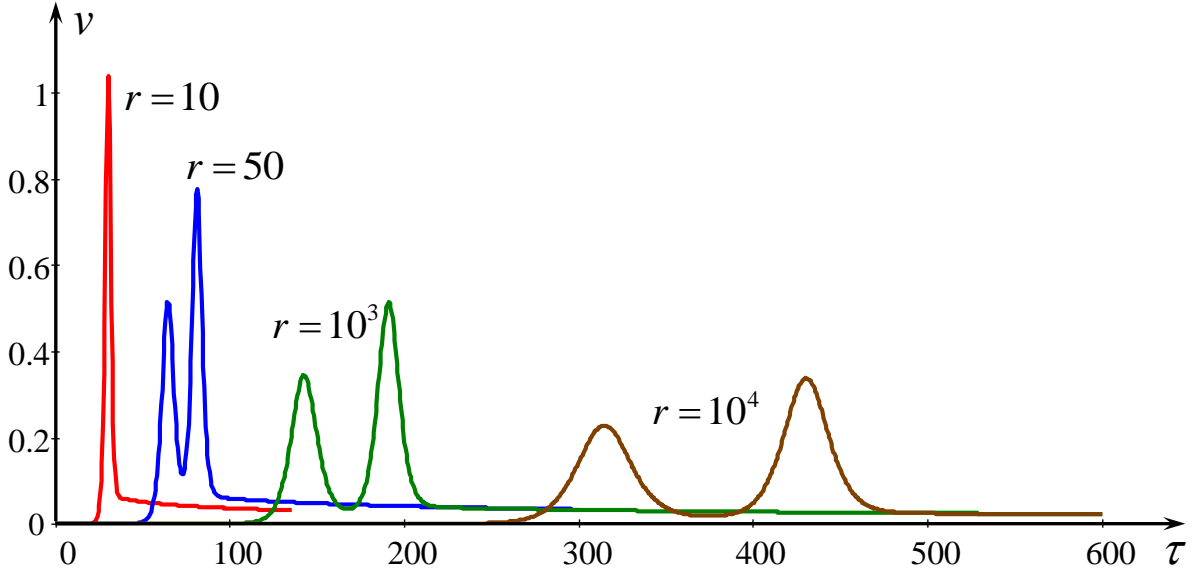
\begin{figure}[h!]
\centering
\includegraphics[width=16cm]{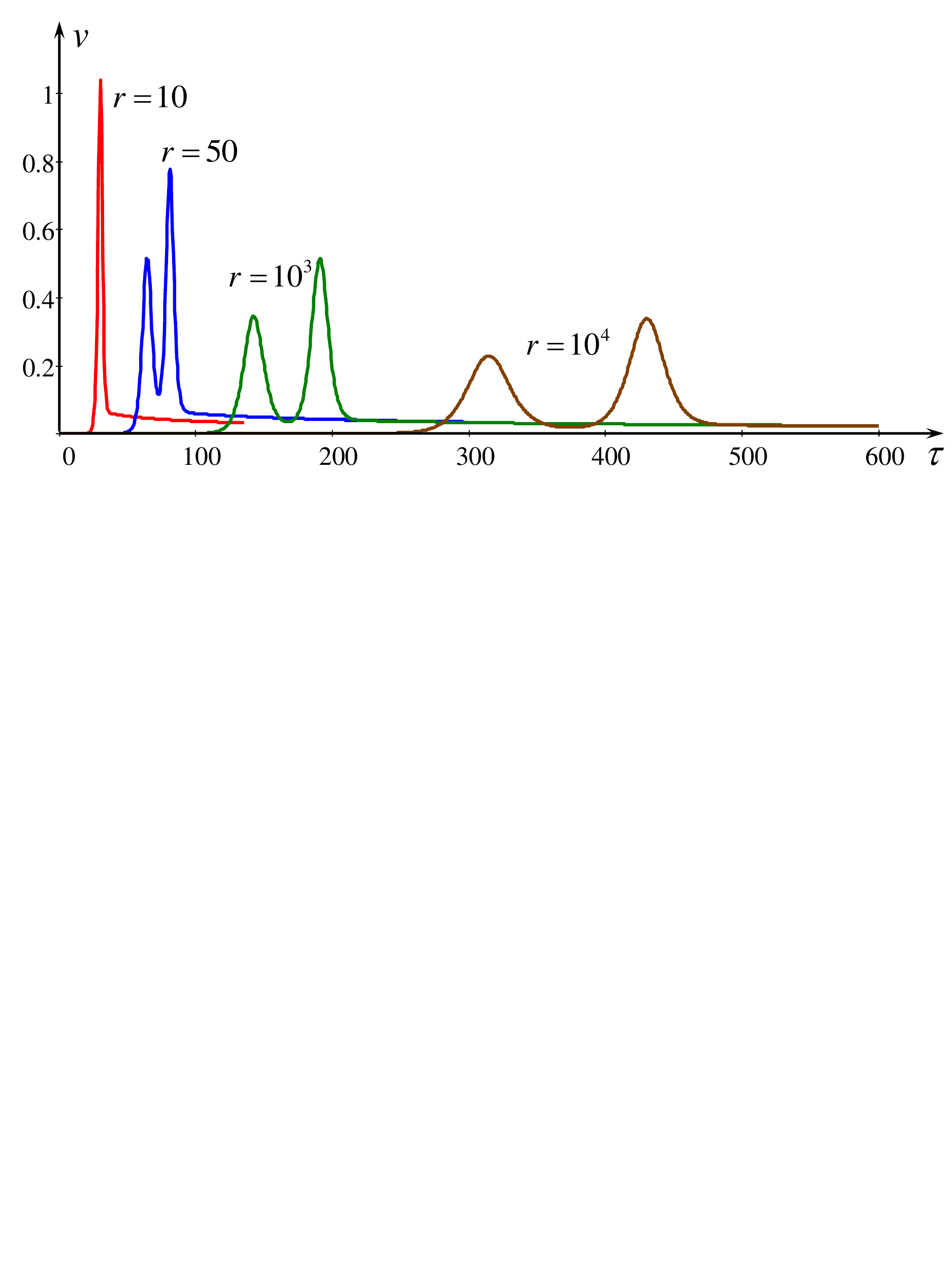}
\begin{picture}(300,6)%
\put(-58,410){$0$}%
\end{picture}
\vspace{-13.5cm}
\caption{(color online). Fission of initial pulse at $r = 10$ onto two ss-solitons within the exact solution described by Eqs. (\ref{eq:005})--(\ref{eq:006b}) with the following parameters: $\varepsilon = 10^{-4}$, $\rho_1 = 0.1$, $\rho_2 = 10^{-4}$,  $\tau_1 = 0$, and $\tau_2 = -10$.
To make graphics clearly visible, we multiplied function $v(\tau)$ by 5 at $r = 50$, by 15 at $r = \cdot 10^3$, and by 45 at $r = 10^4$.} %
\label{Fig07} %
\end{figure}

There is also the process of fission of an initial pulse-type perturbation into ss-solitons that looks very similar to the pure soliton breakdown of a pulse in the plane KdV equation.
An example of such a process is shown in Fig. \ref{Fig07}.

The physical importance of such solutions is not clear but mathematically they are very interesting.
Johnson in his paper \cite{Johnson-1980} mentioned that the ``choice of either Bi or Ai functions does not lead to a proper solution of the cKdV equation'' but he assumed that, perhaps, there is some mileage in describing the evolution of pulse-type initial profiles in terms of such functions.

\section{Pulse disintegration into KdV-like solitons and interaction of KdV solitons with ss-solitons}
\label{Sect3}%

\begin{figure}[h!]
\centering
\includegraphics[width=10cm]{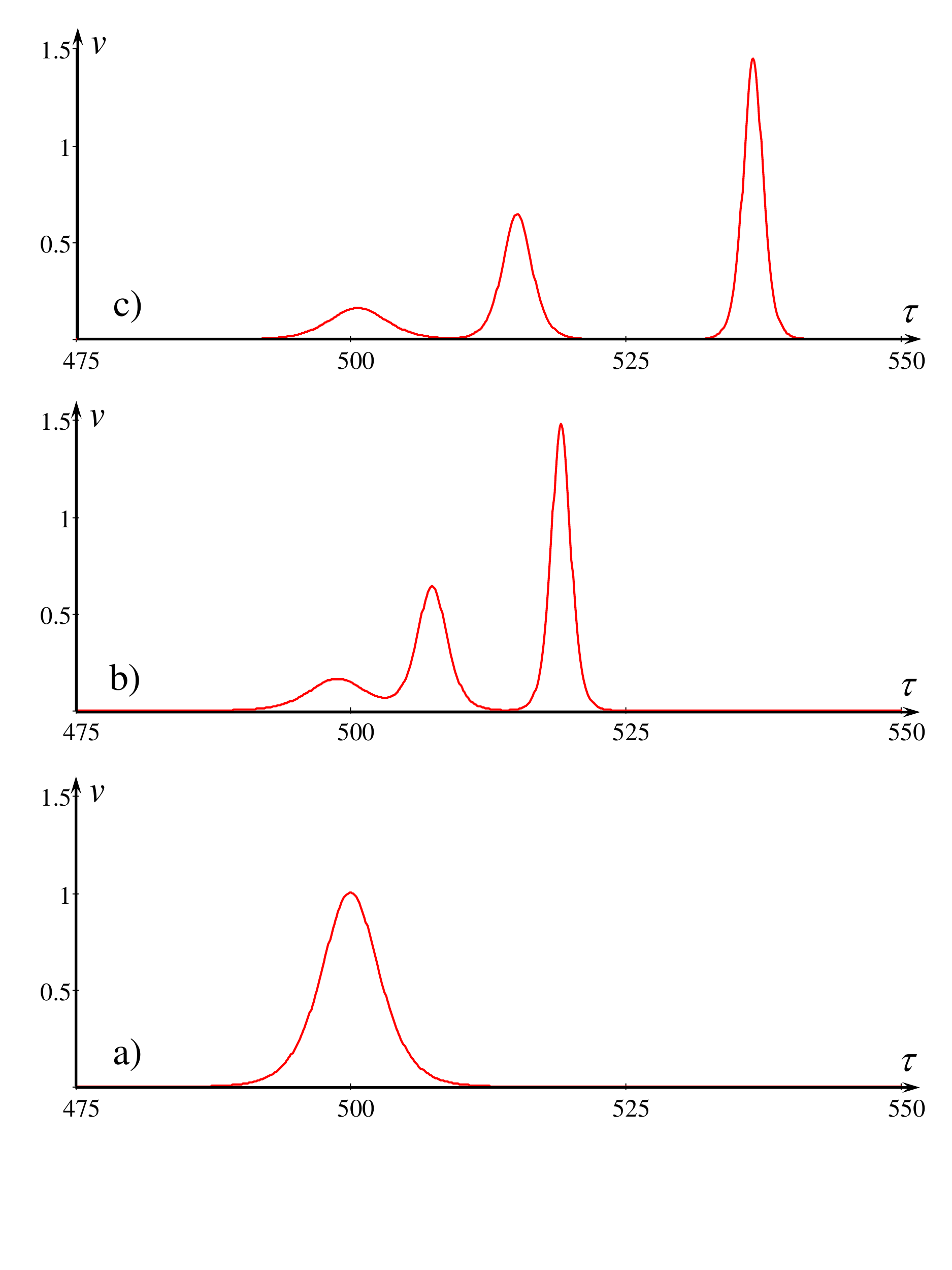}
\begin{picture}(300,6)%
\put(23,65){\small $0$}%
\put(23,178){\small $0$}%
\put(23,289){\small $0$}%
\end{picture}
\vspace{-2.0cm}%
\caption{(color online).
Initial pulse disintegration in the cKdV equation and emergence of KdV-like solitons. 
Frame a) $r - r_0 = 0$, frame b) $r - r_0 = 6$, frame c) $r - r_0 = 12$.}%
\label{Fig08} %
\end{figure}
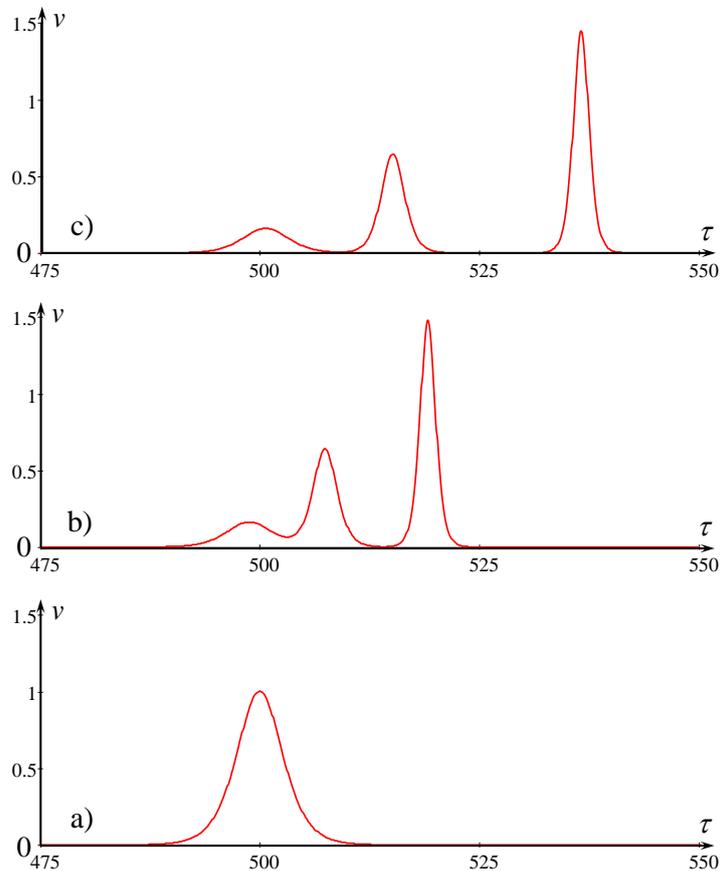

As was shown above, a KdV soliton is very robust in the cylindrical system and keeps its identity even in the process of decay due to geometrical divergence.
The interaction between two KdV-like solitons is very much similar to the interaction of KdV solitons in the plane case.
It is natural to expect that solitons can emerge from wide initial pulses in the same manner as in the plane case.
To confirm this conjecture, we conducted numerical experiments with wider initial pulses which gives rise to the emergence of several solitons in the plane KdV equation.
The typical example with three solitons emergence is shown in Fig. \ref{Fig08}.
This example corresponds to the pure soliton decay of a $\sech^2$-pulse in the plane KdV equation. 
We see that in the cKdV equation the same pure soliton decay occurs at the early stage of evolution and then, each soliton experiences the adiabatic decay in accordance with the asymptotic formulae (\ref{KdV-approx}) and (\ref{Adiabat}).

A similar pulse disintegration into a number of solitons was observed for pulses of positive polarity and different initial duration and amplitudes.
A pure soliton disintegration was observed for the same parameters of an initial pulse as in the plane case.
In general, the initial pulse breaks into solitons and a trailing dispersive wave train. 
Fission into solitons was also observed in a recent paper \cite{Tseluiko-2022}

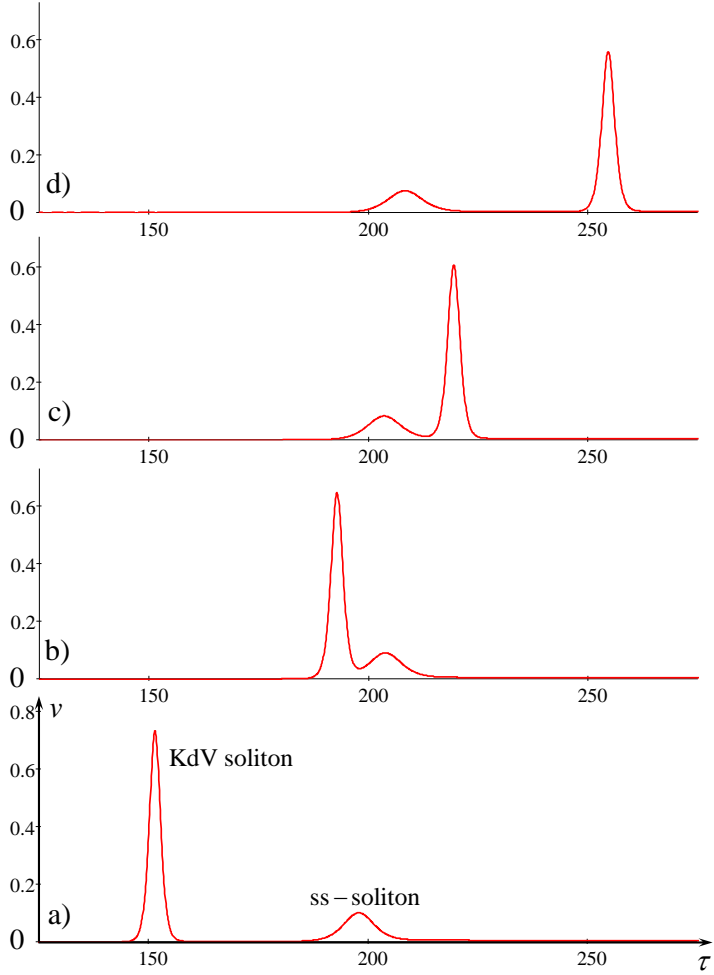
\begin{figure}[h!]
\centering
\includegraphics[width=10cm]{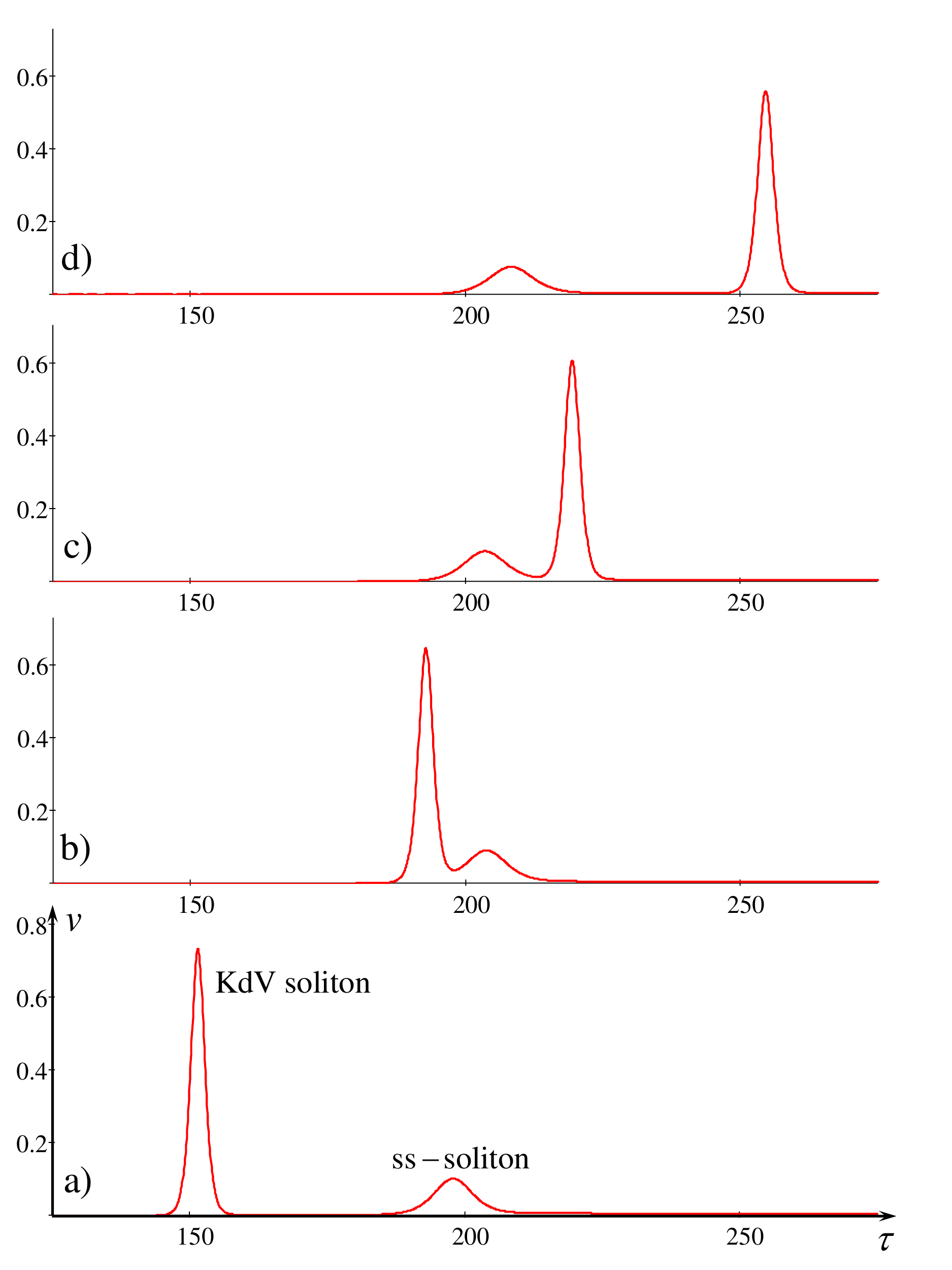}
\begin{picture}(300,6)%
\put(14,28){\small $0$}%
\put(14,127){\small $0$}%
\put(14,217){\small $0$}%
\put(14,303){\small $0$}%
\end{picture}
\vspace{-0.5cm}
\caption{(color online).
Interaction of the KdV-like soliton with the ss-soliton in outgoing cylindrical waves. 
The amplitude of the KdV soliton was $A_0 = 1$ at $r_0 = 100$.
The parameters of the ss-solton were $\varepsilon = 10^{-10}$, $\rho = 1$, $\tau_1 = 1$. 
Frame a) $r - r_0 = 160$, frame b) $r - r_0 = 190$, frame c) $r - r_0 = 210$, frame d) $r - r_0 = 240$.}%
\label{Fig09} %
\end{figure}

It is of interest to study also the interaction of a KdV soliton with an ss-soliton.
This can be done numerically for the initial condition consisting of one KdV soliton and one ss-soliton.
The result of such interaction is shown in Fig. \ref{Fig09}.

Thus, we see that the traveling KdV-type soliton overtakes the ss-soliton and after the interaction, both of them restore their shapes and continue moving and decaying due to the geometrical divergence.
Thus, we can conclude that in the weakly nonlinear physical systems with a small dispersion, the outgoing pulses with cylindrical fronts evolve in a similar way as in the plane KdV equation but experience amplitude decay due to the geometrical divergence.

\section{Concluding remarks}
\label{Sect4}%

In this paper, we have presented a detailed analysis of solitary wave solutions to the cylindrical KdV equation. 
It was shown that soliton-like solutions in the form of KdV solitons exist in this equation. 
In the process of geometrical divergence, such solitons gradually decay so that the total energy of the initial pulse is conserved, $E = \int \eta^2r\,d\tau = \mbox{const}$, where the integration should be carried out over $\tau$ in the infinite limits, $-\infty < \tau < +\infty$.
There are also exact solutions of the cKdV equation \cite{Nakamura-1981b} which have pulse-type shapes (ss-solitons) which are very similar to KdV solitons of the same amplitudes.
Their parameters (amplitudes and duration) vary with the distance in the same manner as in the diverging KdV-like solitons, $A \sim r^{-2/3}$, $T \sim r^{1/3}$.
However, such solutions are not traveling waves but are closer to self-similar solutions.

A numerical study of interactions between KdV-like solitons, ss-solitons, as well as between KdV-like and ss-solitons revealed that all of such solitons are robust and, apparently, interact elastically.
A general pulse-type initial perturbation of positive polarity in the course of evolution experiences a breakdown into a number of KdV-like solitons and trailing dispersive wavetrain.
Each of emerged KdV-like solitons decays then individually due to the geometrical divergence. 

In conclusion, we note that some asymptotic solutions to the cKdV equation were obtained in Refs. \cite{Santini-1979, Santini-1980}.
Using symbolic computation, Gao and Tian \cite{Gao-1999} constructed a few self-similar solutions to the cKdV equation; some of them were mentioned in this paper and obtained by other authors using analytical methods. 
However, all these solutions are out of our current interest as they are not of a soliton-type.

In perspective, we plan to study quasi-cylindrical waves within the cylindrical version of the Kadomtsev--Petviashvili equation (alias Johnson equation) \cite{Johnson-1980}. 
The important problem to be studied is the stability of a soliton front with respect to small azimuthal perturbations and lump formations.
One more problem to be studied in perspective is the dynamics of solitons within the cylindrical Gardner equation containing both quadratic and cubic nonlinearities. 
Such an equation is applicable to the description of internal waves in the ocean and the results obtained can be of practical interest.
\\

\noindent {\it Acknowledgements.} W.H. acknowledges the financial support from China Scholarship Council (grant No. 202002425001). His study was also supported by the National Natural Science Foundation of China (grants No. 11947093 and 12204554) and the Natural Science Foundation of Henan Province of China (grant No. 222300420393).
Y.S. acknowledges the financial support provided by the President Council of the Russian Federation (grant No. NSH-70.2022.1.5) for the State support of Leading Scientific Schools of the Russian Federation. 
The authors are grateful to K. Khusnutdinova for her helpful remarks.

\end{document}